\documentclass{mem}
\usepackage{tx fonts}\usepackage{balance}
\usepackage{graphicx}
\usepackage[a4paper,breaklinks,dvipdfm]{hyperref}
\idline{0}{0}
\begin{document}
\def\teff{$T\rm_{eff }$}
\def\kms{$\mathrm {km s}^{-1}$}

\title{
Extremely dark GRBs: the case of GRB\,100614A and GRB\,100615A}

   \subtitle{}

\author{
V. \,D'Elia\inst{1,2} 
\and G. \,Stratta\inst{1,2}
          }

  \offprints{V. D'Elia}

\institute{
ASI-Science Data Center, Via Galileo Galilei, I-00044 Frascati, Italy
\and
INAF-Osservatorio Astronomico di Roma, Via Frascati 33, I-00040 Monteporzio Catone, Italy
\email{delia@asdc.asi.it}
}

\authorrunning{D'Elia \& Stratta }

\titlerunning{Extremely dark GRB\,100614A and 15A}

\abstract{ 

  Dark gamma-ray bursts (GRBs) are sources with a low optical-to-X-ray
  flux ratio. Proposed explanations for this darkness are: i) the GRB
  is at high redshift ii) dust in the GRB host galaxy absorbs the
  optical/NIR flux iii) GRBs have an intrinsically faint afterglow
  emission. Within this framework, GRB\,100614A and GRB\,100615A are
  extreme. In fact, they are bright in the X-rays, but no optical/NIR
  afterglow has been detected for either source, despite several
  follow-up campaigns began early after the triggers.  We analyze the
  X-ray data and collect all the optical/NIR upper limits in
  literature for these bursts. We then build optical-to-X-ray spectral
  energy distributions (SEDs) at the times at which the reddest upper
  limits are available, and we model our SEDs with the extinction
  curves of the Milky Way (MW), Small Magellanic Cloud (SMC), and the
  attenuation curve obtained for a sample of starburst galaxies.  We
  find that to explain the deepest NIR upper limits assuming either a
  MW or SMC extinction law, a visual extinction of $A_V > 50$ is
  required, which is extremely unlikely. Since both GRBs are bright in
  X-rays, explanation iii) also cannot explain their dark
  classification, unless optical radiation and X-rays are not part of
  the same synchrotron spectrum. An alternative, or complementary
  explanation of the previous possibility, involves greyer extinction
  laws. A starburst attenuation curve gives $A_V>10$, which is less
  extreme, despite still very high. Assuming high redshift in addition
  to extinction, implies an $A_V>10$ at $z=2$ and $A_V>4-5$ at $z=5$,
  regardless of the adopted extinction recipe. A different, exotic
  possibility would be an extremely high redshift origin ($z>17$ given
  the missing K detections). Population III stars are expected to
  emerge at $z \sim 20$ and can produce GRBs with energies well above
  those inferred for our GRBs at these redshifts.  Mid- and far-IR
  observations of these extreme class of GRBs can help us to
  differentiate between the proposed scenarios.   }
\maketitle{}

\section{Introduction}

Long-duration gamma-ray bursts (GRBs) are high energy phenomena linked
to the death of massive stars, emitting most of their radiation in the
hundreds of keV range.  The gamma-ray (or prompt) event is followed by
an afterglow at longer wavelengths, which has proven crucial to
understand the physics of these sources and to investigate the nature
of their surrounding medium.

While the X-ray (0.1-10 keV) afterglows are detected in virtually all
GRBs, the optical/near infrared (NIR) counterparts are more elusive.
Just a few months after the discovery of the first optical counterpart
of a GRB (van Paradijs et al. 1997), the search for an optical
afterglow associated to GRB 970828 was unsuccessful (Groot et
al. 1998) leading to the definition of `dark burst' for a GRB with an
X-ray counterpart, but not an optical one. Initially, this lack of
detection was widely ascribed to the delay time between the GRB event
and optical observations, since ground-based facilities could be on
target only hours after the trigger, when the afterglow had faded
below their sensitivity limit.

The main scientific driver of the {\it Swift} satellite was to
facilitate the GRB afterglow detection by quickly repointing its
narrow field instruments and disseminating the GRB coordinates
worldwide. The quick detection with XRT and the increasing number of
ground-based automated facilities dramatically improved the optical
follow-up success, reducing the fraction of dark GRBs. The definition
of dark GRB was reconsidered in a more quantitative manner, based on
the expected spectral behaviour in the context of the most commonly
accepted scenario of the fireball model. Van der Horst et al. (2009)
propose to classify a dark GRB using only the spectral indices of the
power-law slopes in the optical ($\beta_O$) and X-ray ($\beta_X$)
data, where $\beta$ is linked to the flux by the relation
$F_{\nu}\propto \nu^{-\beta}$. They noted that, regardless of many
assumptions on the specific electron energy distribution, if both the
optical and X-ray radiation are produced by synchrotron emission from
the same source, $\beta_O$ and $\beta_X$ are linked. In particular,
$\beta_O=\beta_X -0.5$ if the cooling frequency lies between the
optical and the X-rays, and $\beta_O=\beta_X$ otherwise. Thus, the
optical-to-X-ray spectral index allowed range is
$\beta_X-0.5\le\beta_{OX}\le\beta_X$, with $\beta_{OX}=\beta_X-0.5$ if
a spectral break is present just below the lowest X-ray energy
detected. GRB afterglows with $\beta_{OX}<\beta_X-0.5$ are classified
as dark in this picture.

The optical darkness can be ascribed to different factors (see e.g.,
Perley et al. 2009): i) the GRB can be at high
redshift, so that the Lyman-$\alpha$ absorption prevents optical
identifications; ii) dust in the GRB host galaxy or along the line of
sight can absorb the optical afterglow counterpart; iii) the optical
faintness can have an intrinsic origin. In this case, however, also
the X-ray counterpart is expected to be faint.

Here we study the `darkness' properties of two GRBs (namely,
GRB\,100614A and GRB\,100615A), which are very bright in X-rays, but
are not detected in the optical/near-IR band and have no reported host
galaxy candidate. The next section summarizes the observations of
these two sources and our analysis method. Sect. 3 illustrates our
results, and in Sect. 4 we discuss our findings and draw our
conclusions. Decay, photon, and spectral indices are indicated with
$\alpha$, $\Gamma$, and $\beta$, following the standard convention
$t^{-\alpha}$, $N_{ph}^{-\Gamma}$, and $\nu^{-\beta}$, respectively.

\begin{figure}[t!]
\resizebox{\hsize}{!}{\includegraphics[clip=true]{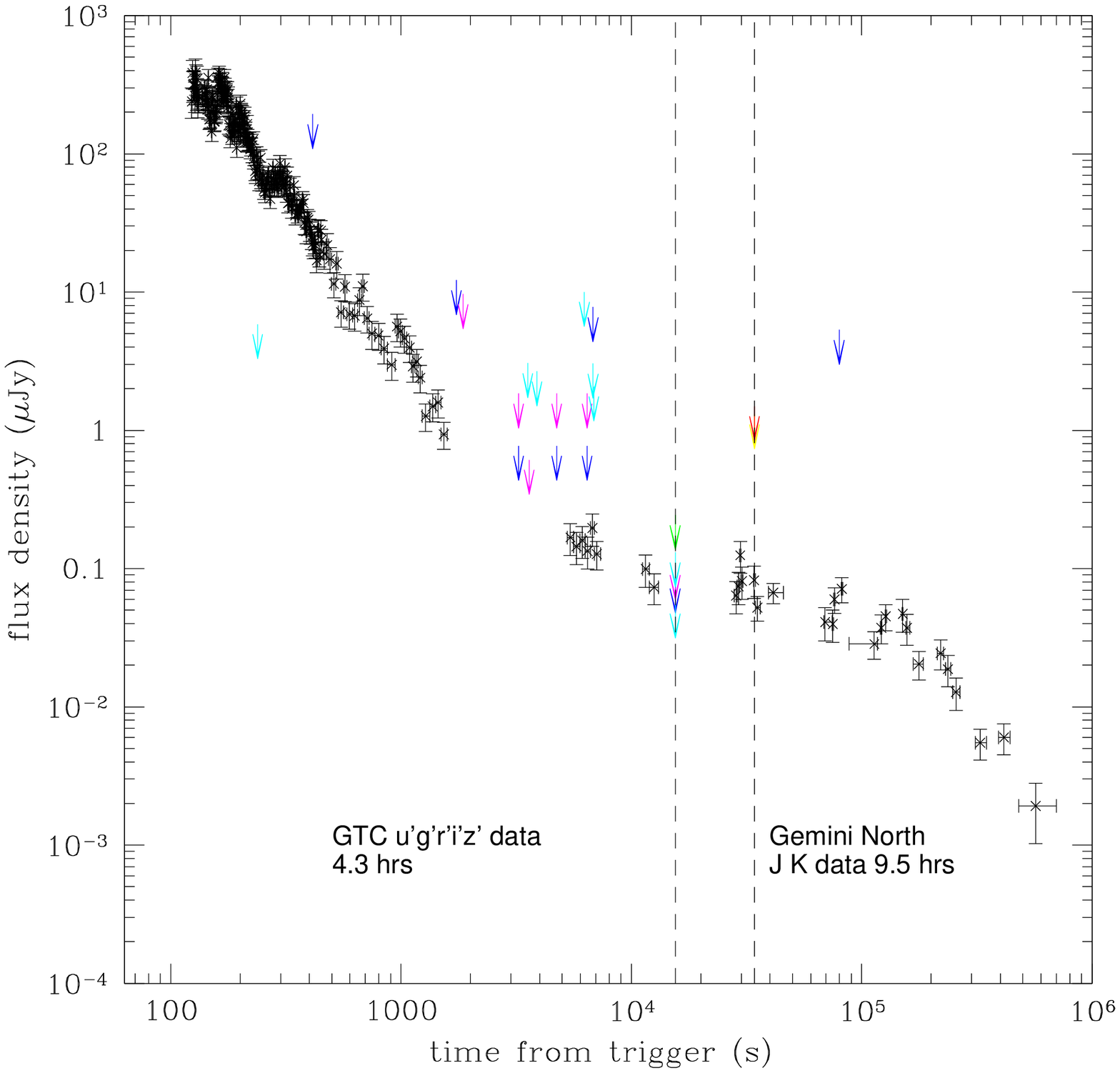}}
\resizebox{\hsize}{!}{\includegraphics[clip=true]{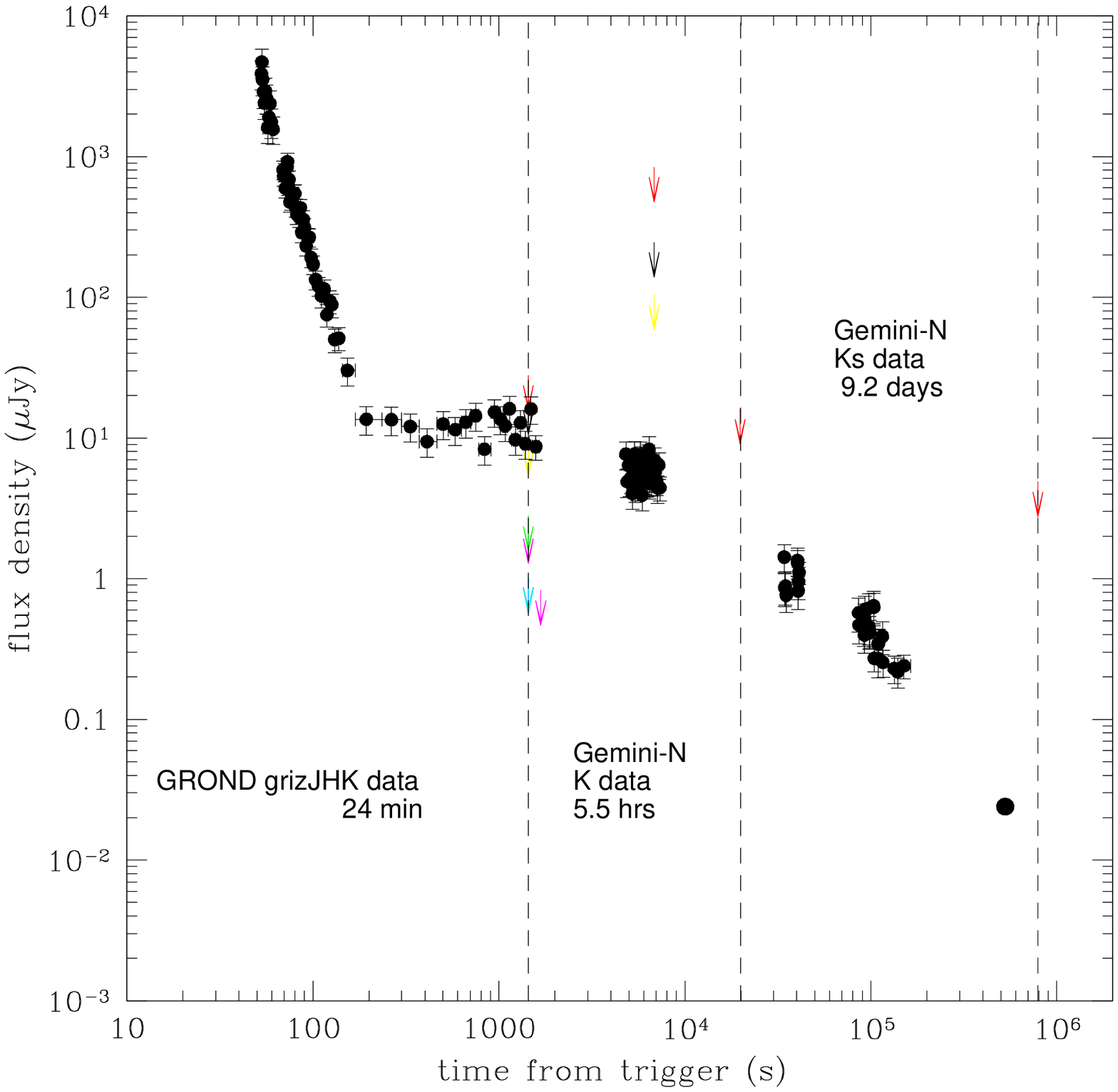}}
\caption{\footnotesize The GRB\,100614A (upper panel) and GRB\,100615A
  (bottom panel) X-ray lightcurves, together with all the optical/NIR
  upper limits estimated using several ground-based facilities (see
  D'Elia \& Stratta 2011 for details). Cyan, blue, magenta, green,
  yellow, black and red upper limits represents the g' (and all those
  bands bluer than g'), R, I, z, J, H, K bands, respectively. Vertical
  dashed lines emphasize the reddest observations and the time at
  which they have been obtained. Such observations are used in our
  analysis. The faintest X-ray data in the bottom panel draws the
  GRB\,100615A {\it Chandra} observation.}
\end{figure}

\section{Observations and analysis}

GRB\,100614A and 15A were both discovered by {\it Swift}-BAT and
immediately repointed by XRT and UVOT. XRT reported an extremely
bright transient in both cases, while UVOT did not detect any optical
counterpart. The two sources were followed-up by several ground-based
facilities, beginning 4 hr and 24 minutes after the trigger,
respectively, but no optical/NIR counterpart down to the K band has
been detected. GRB\,100615A has been observed even 9.2 days after the
trigger in a deep, K band exposure, but again no counterpart has been
found. All the data collected for these two bursts are plotted in
Fig. 1. We refer the reader to D'Elia \& Stratta (2011) for a full
description of the observations of these GRBs and for a comprehensive
reference list.

\subsection{Analysis}

We extract the optical to X-ray spectral energy distribution (SED) for
both GRBs by selecting those epochs at which we have the deepest and
reddest observations (i.e. less affected by any dust extinction), to
constrain at best the intrinsic optical afterglow flux upper limit.
In addition, we attempt to select an epoch not too close to the
initial X-ray steep decay, which is thought to be produced by a
different component than the one responsible for the afterglow
emission. Magnitudes reported in literature were corrected for
Galactic absorption (E(B-V) = $0.03$ and $0.047$ for GRB\,100614A and
15A, respectively).

A broken power-law model was fitted to the XRT data following the van
der Horst et al. (2009) method, therefore fixing the SED normalization
and the high-energy spectral index to the value obtained from our
X-ray data analysis (within its $90\%$ confidence range), the spectral
break at the X-ray energies and the optical to X-ray spectral index as
$\beta_{OX}=\beta_X-0.5$.

We model the optical suppression from the X-ray extrapolation assuming
either a Milky Way (MW) or a Small Magellanic Cloud (SMC) extinction
curve.  We also test the attenuation curve obtained for a sample of
starburst galaxies (Calzetti et al. 1994). We consider the upper
limits as positive detections, and we verify that the model-predicted
fluxes are consistent with the data, i.e. equal to or below the upper
limits.

\begin{table}[ht]
\begin{center}
  \caption{Lower limits to the visual extinction towards GRB\,100614A and GRB\,100615A.} 
{\footnotesize \smallskip
\begin{tabular}{|lcc|ccc|}
  \hline 
  GRB                &$t-T$  & $\beta_X$& $A_{V,MW}$ & $A_{V,SMC}$    & $A_{V,SB}$    \\
  		     & min   &          &  mag      & mag            & mag          \\
  \hline
  14A       & $258$         & $1.20$    &$8$         & $8$     & $6$        \\
  {\bf 14A} & {\bf 258}     & {\bf 1.50}&{\bf 13}    & {\bf 13}& {\bf 8}     \\
  14A       & $258$         & $1.80$    &$18$        & $18$    & $11$       \\
  14A       & $570$         & $1.20$    &$25$        & $31$    & $6$        \\
  {\bf 14A} & {\bf 570}     & {\bf 1.50}&{\bf 47}    & {\bf 58}& {\bf 11}    \\
  14A       & $570$         & $1.80$    &$69$        & $85$    & $17$       \\
  \hline                                
  15A       & $24$          & $1.20$    &$39$        & $47$    & $10$       \\
{\bf 15A} & {\bf 24}      &{\bf 1.35}   &{\bf 49}    & {\bf 60}& {\bf 13}    \\
  15A       & $24$          & $1.50$    &$59$        & $73$    & $16$       \\
  15A       & $330$         & $1.20$    &$47$        & $58$    & $12$       \\
{\bf 15A}  & {\bf330}      & {\bf 1.35} &{\bf 58}    & {\bf 72}& {\bf 15}    \\
  15A       & $330$         & $1.50$    &$69$        & $86$    & $18$       \\
  15A       & $13200$       & $1.20$    &$12$        & $15$    & $3$        \\
{\bf 15A}  & {\bf13200}& {\bf 1.35}     &{\bf 22}    & {\bf 27}& {\bf 6}     \\
  15A       & $13200$       & $1.50$    &$33$        & $40$    & $9$        \\
  \hline
\end{tabular}
}
\end{center}
\end{table}

\begin{figure}
\centering
\includegraphics[angle=-0,width=3.23cm]{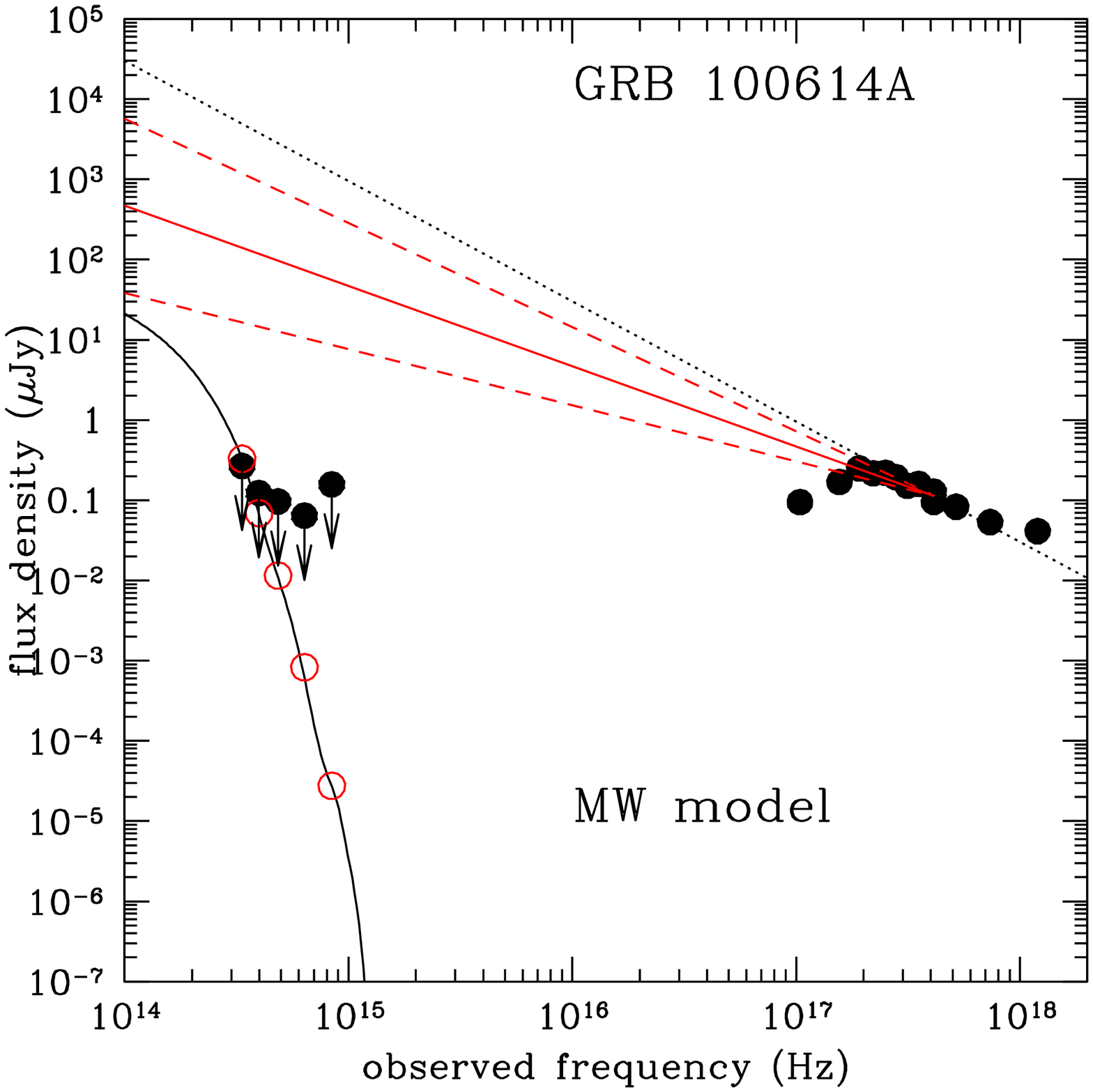}
\includegraphics[angle=-0,width=3.23cm]{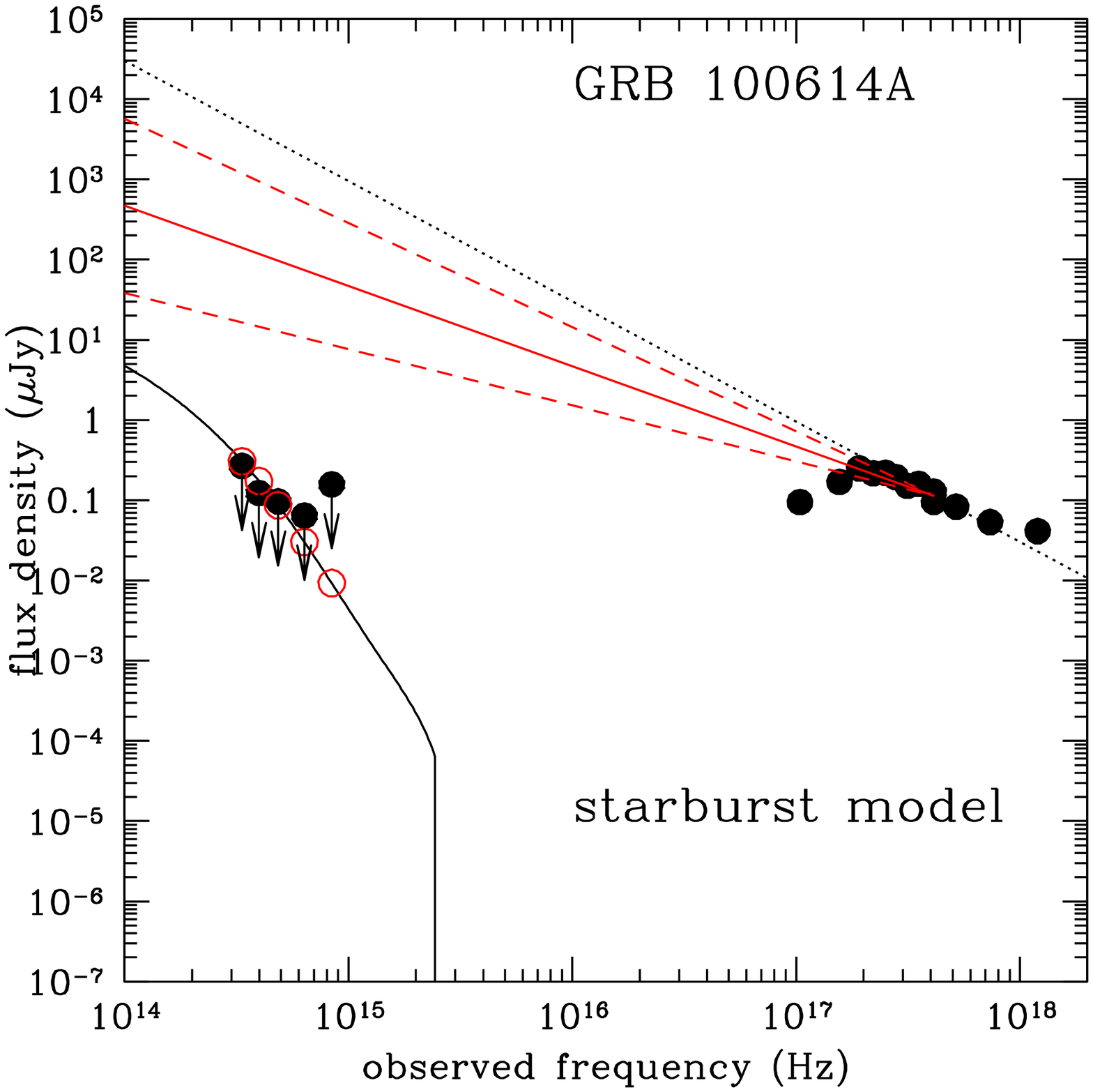}

\includegraphics[angle=-0,width=3.23cm]{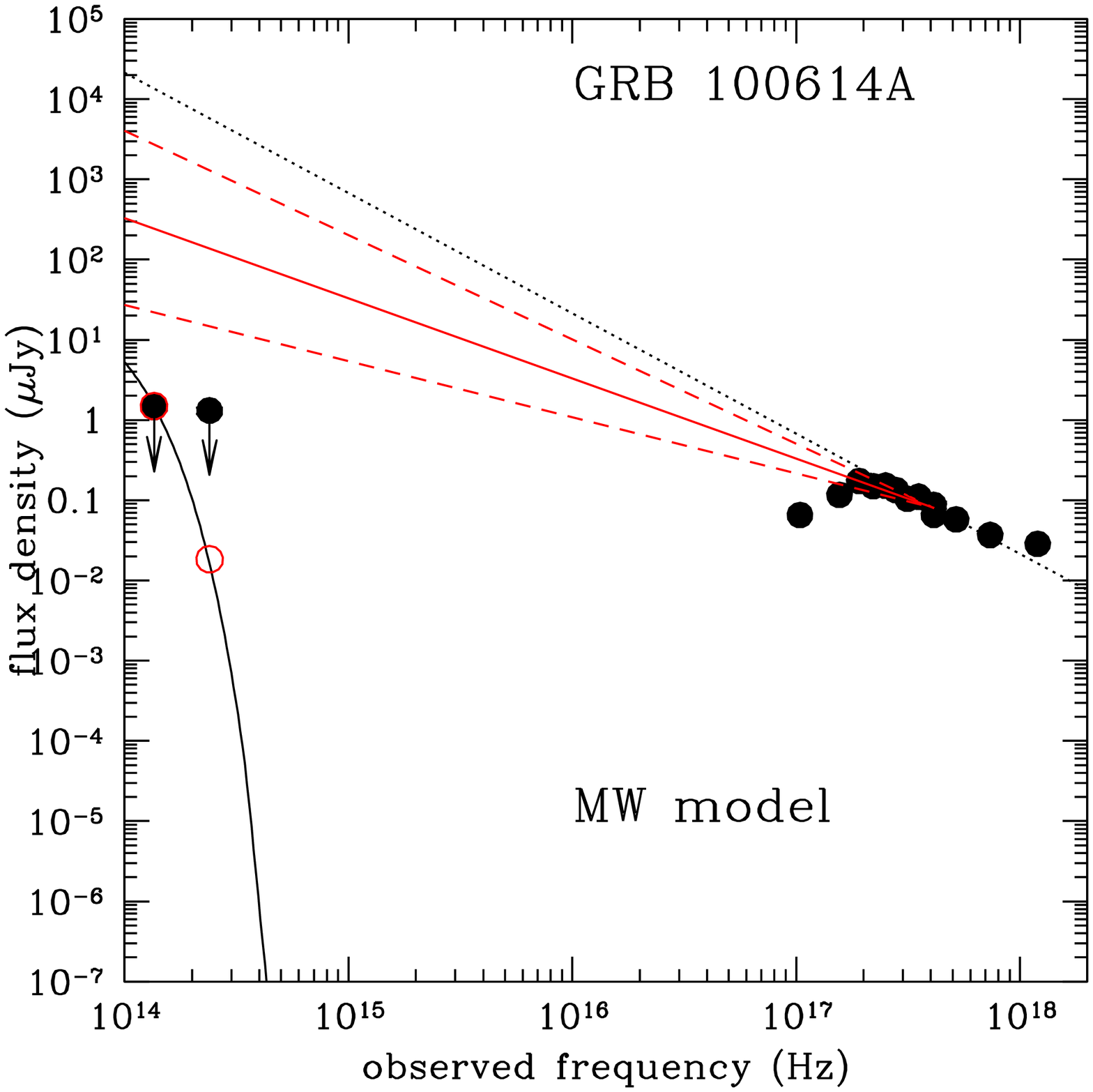}
\includegraphics[angle=-0,width=3.23cm]{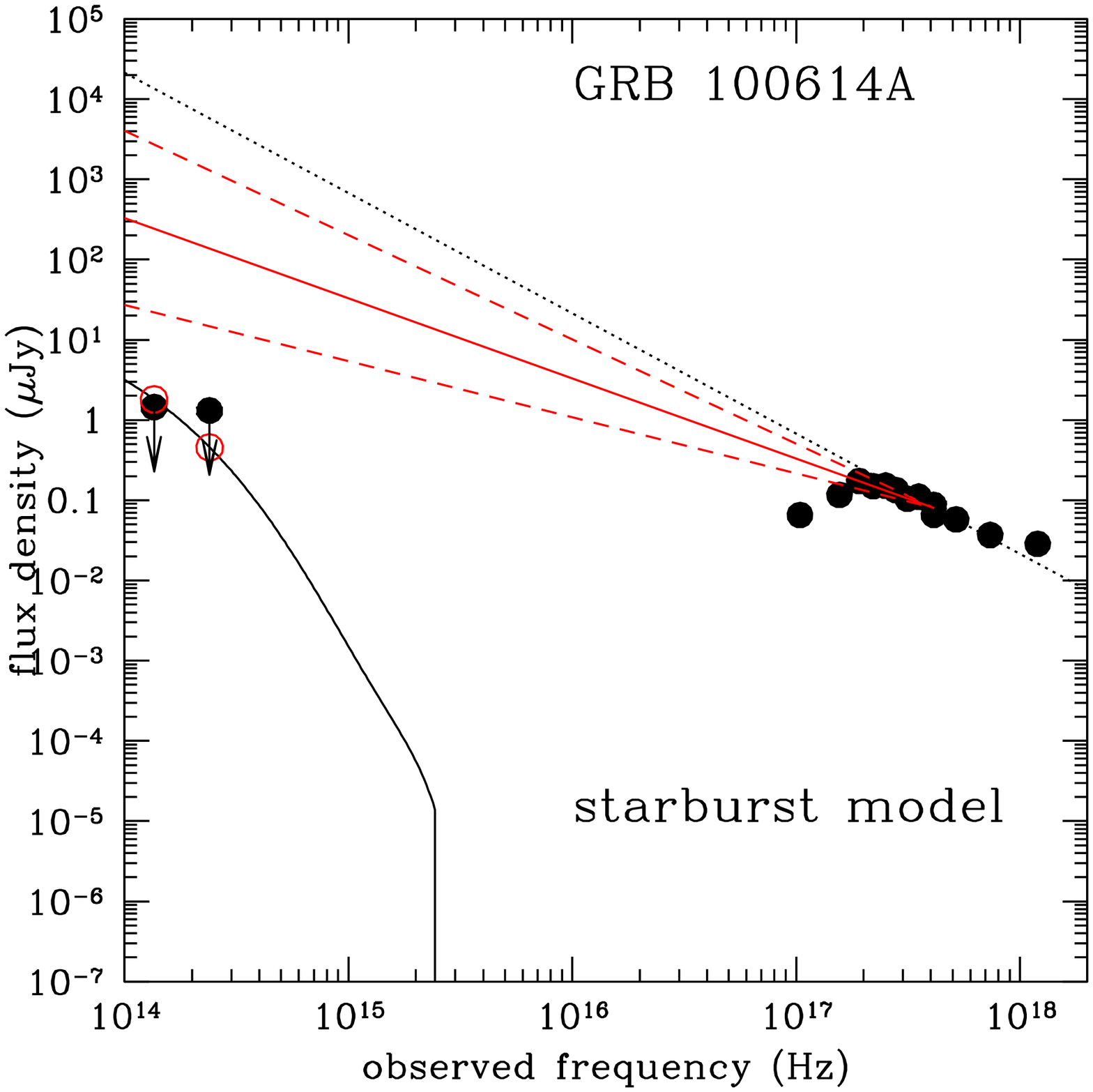}
\caption{GRB\,100614A SEDs at two different epochs: GTC $ugriz$ upper
  limits taken 4.3 hours after the trigger (upper panels) and the NIR
  J and K band taken 9.5 hours after the trigger (lower panels), with
  $\beta_{OX}$ fixed at $\beta_X-0.5$ (solid red line) and the optical
  suppression modelled with the MW or starburst extinction
  curves. Black dotted line represents the best-fit value for the
  X-ray spectral slope ($\beta_X=1.50$). Dashed red lines enclose the
  $90\%$ uncertainty in $\beta_{OX}$. }
\label{spe1}
\end{figure}

\section{Results}

We find that GRB fluxes computed from both optical and NIR data are
well below the most conservative extrapolation from X-rays and require
strong absorption using the data taken within 1 day after the
trigger. For GRB\,100615A, very late-time data are available (9.2 days
after the trigger), for which the NIR flux is still below the X-ray
extrapolation.

The GRB\,100614A SED with the reddest flux upper limit (i.e. 570
minutes after the trigger) requires a rest-frame V-band dust
extinction of $A_V\ge47$, $58$, and $11$ mag, assuming a MW, SMC, or a
starburst attenuation curve, respectively, and using the best-fit value
for $\beta_X$ (Fig. 2).  Even fixing the optical to X-ray energy
spectral index to its lowest allowed value (within its $90\%$
confidence range), that is, in the most conservative case, results
still provide very high $A_V$ lower limits (Table 1). Less stringent
constraints on $A_V$ are obtained using the GTC $ugriz$ flux upper
limits obtained 258 minutes after the trigger. We obtain $A_V\ge13$
mag with either the MW and SMC extinction curve and $A_V\ge8$ mag with
the starburst case.

For GRB\,100615A, we obtain even tighter lower limits. The SED
extracted 24 minutes after the burst requires $A_V\ge64$, $79$, or
$16$ mag assuming either a MW or SMC extinction curve, or a starburst
attenuation curve, respectively.  These lower limits are still very
high for the SED extracted $5.5$ hours post burst: $A_V\ge58$, $72$,
and $15$ mag for the three extinction recipes. Less critical but still
high values, are obtained even $9.2$ days from the burst: $A_V\ge22$,
$27$, and $6$ mag (Fig.3).  The latter lower limits are lower than the
ones obtained at earlier epochs (but still extreme), possibly due to a
selection effect, since at later times the X-ray flux decreases, but
the optical/NIR upper limits can not become fainter consistently,
owing to the instrument detection limits.

Table 1 reports all the $A_V$ lower limits evaluated from the
GRB\,100614A and GRB\,100615A data at the given mean observation
epochs. These lower limits are computed for the reported ranges of the
X-ray spectral index ($\beta_X=\Gamma-1$, estimated in Sect. 3) that
corresponds to the minimum, maximum, and mean value of the estimated
$90\%$ confidence range. Bold face characters indicate the results
obtained with the most probable X-ray spectral index. Upper limits to
the optical-to-X-ray spectral indices $\beta_{OX}$ are also shown.

\begin{figure}
\centering
\includegraphics[angle=-0,width=3.23cm]{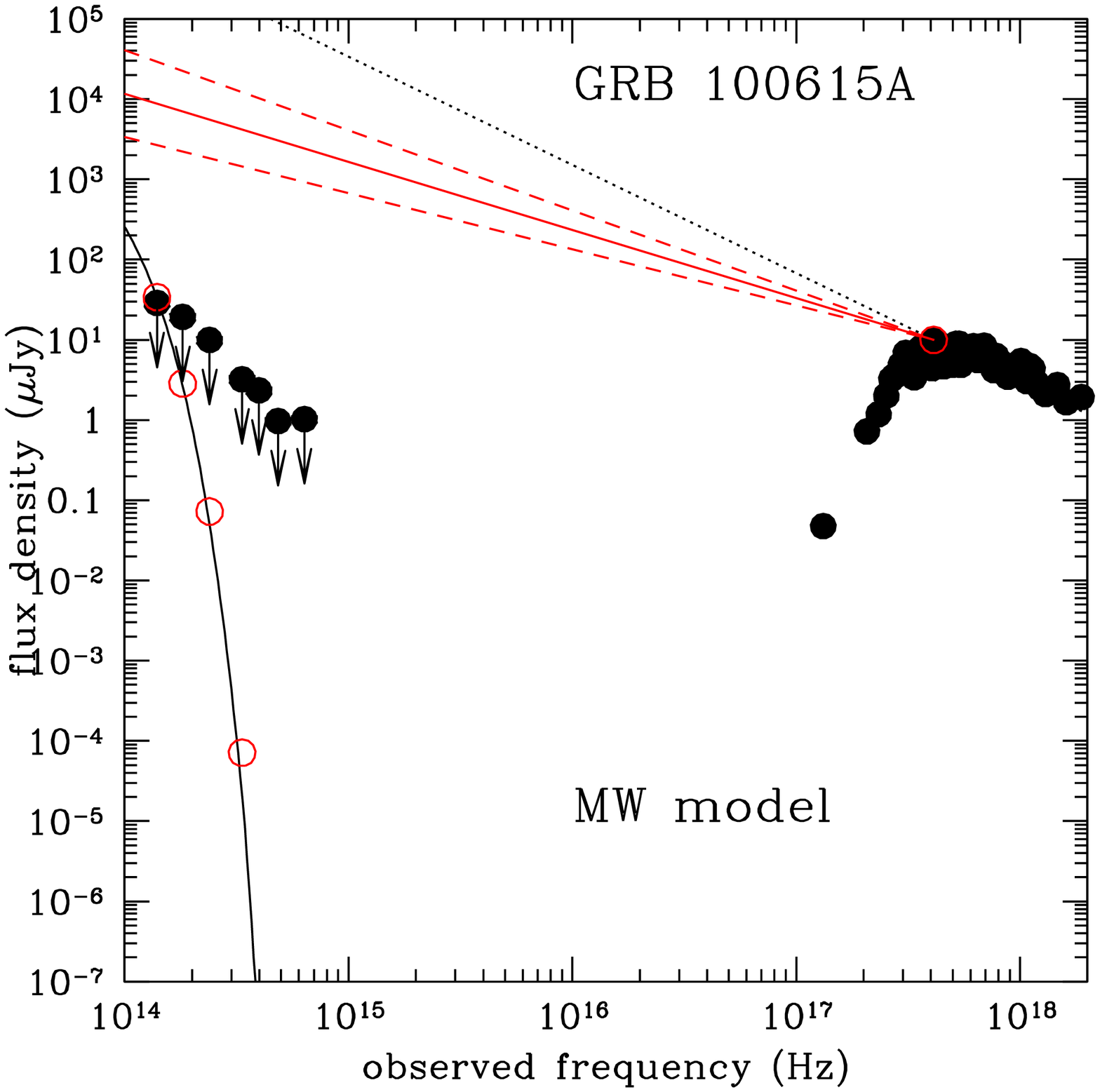}
\includegraphics[angle=-0,width=3.23cm]{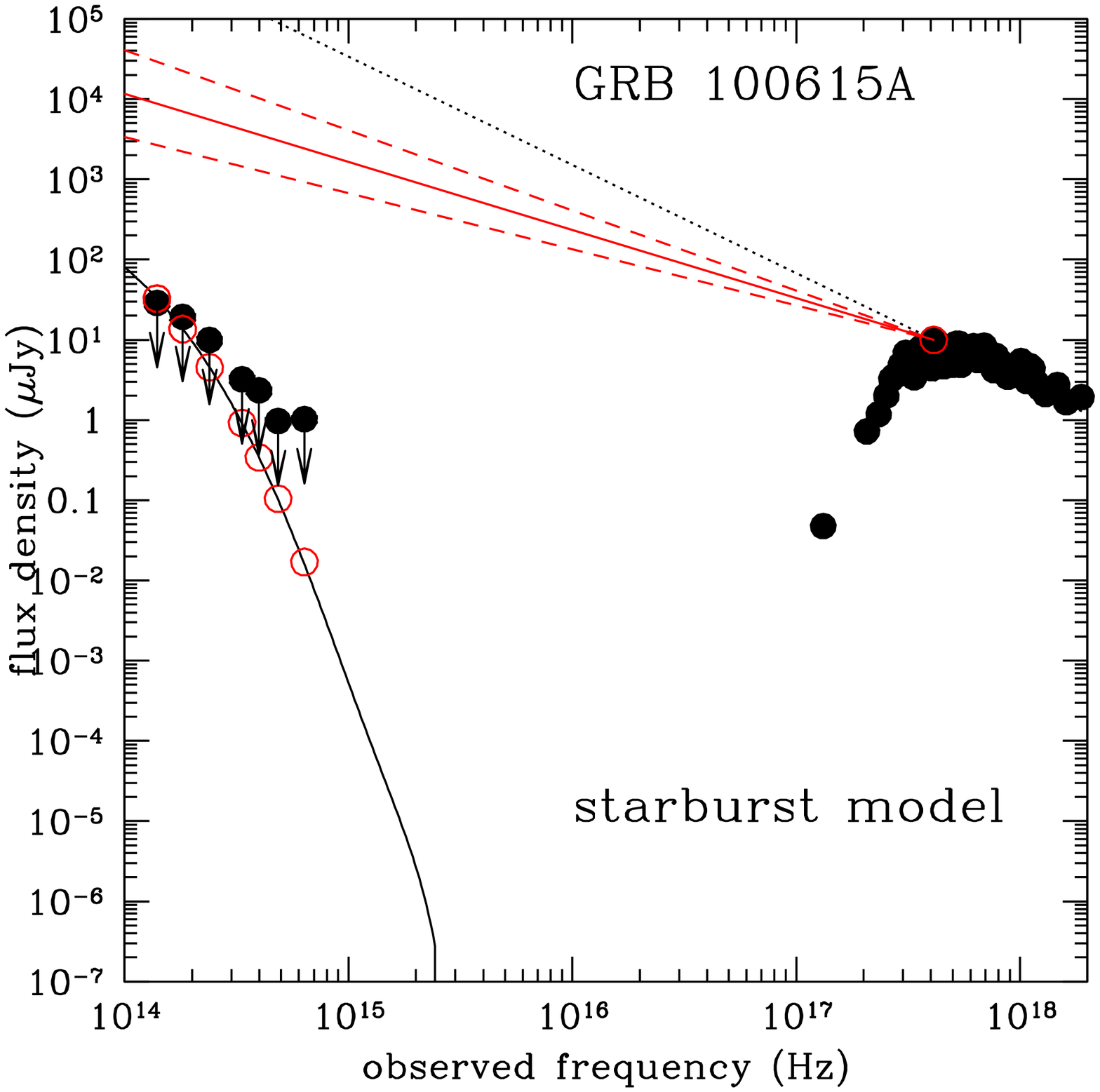}

\includegraphics[angle=-0,width=3.23cm]{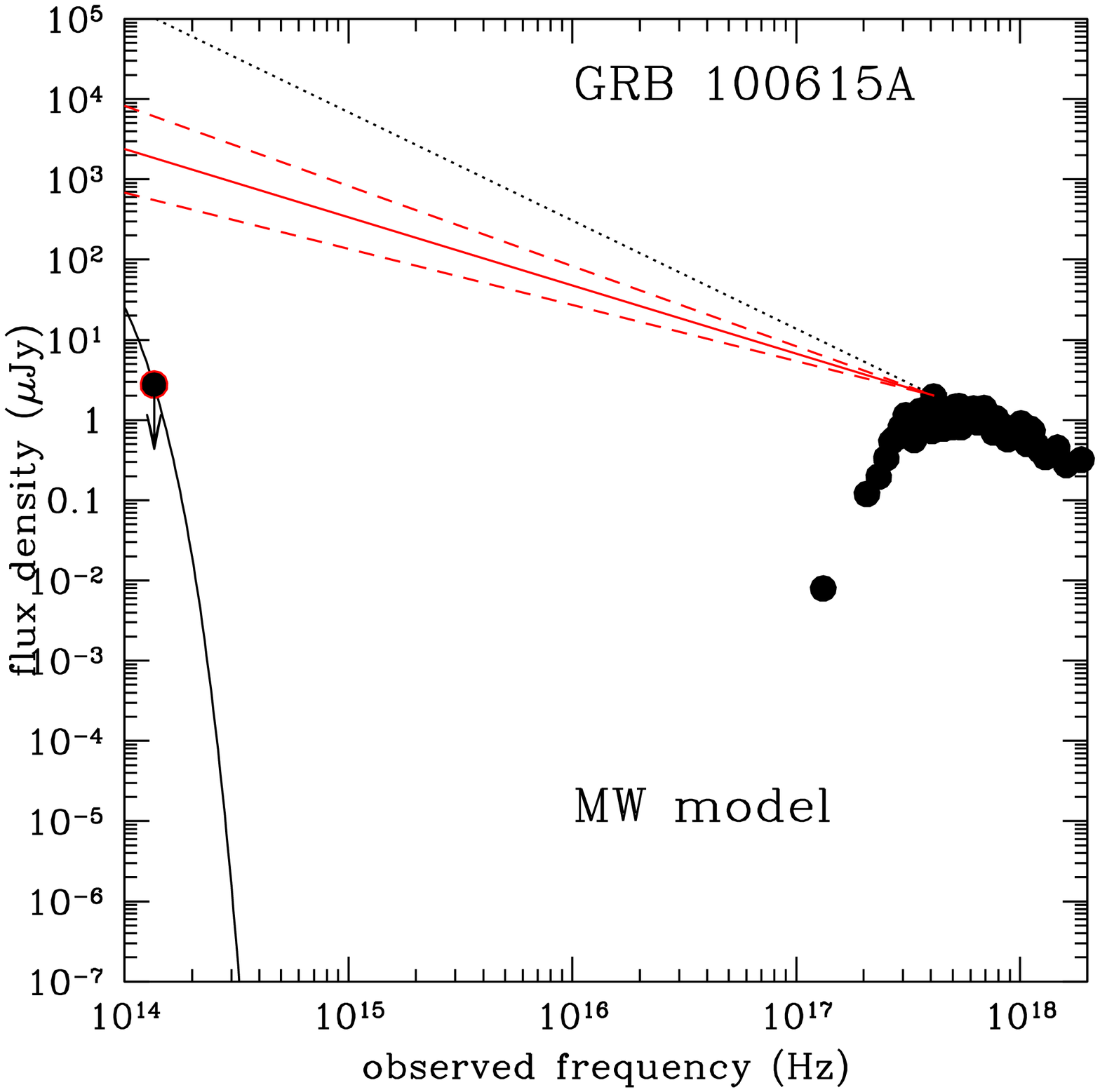}
\includegraphics[angle=-0,width=3.23cm]{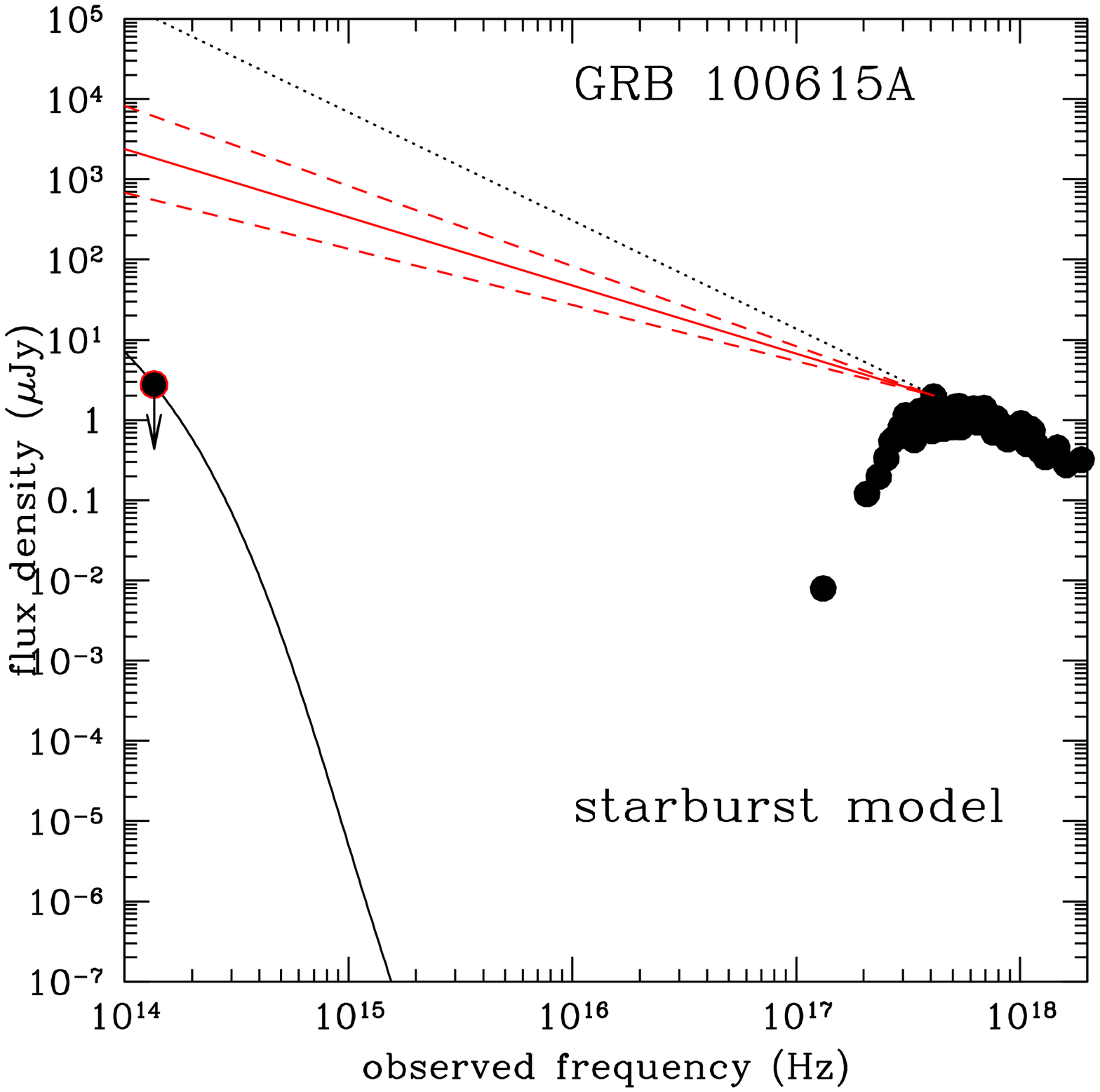}

\includegraphics[angle=-0,width=3.23cm]{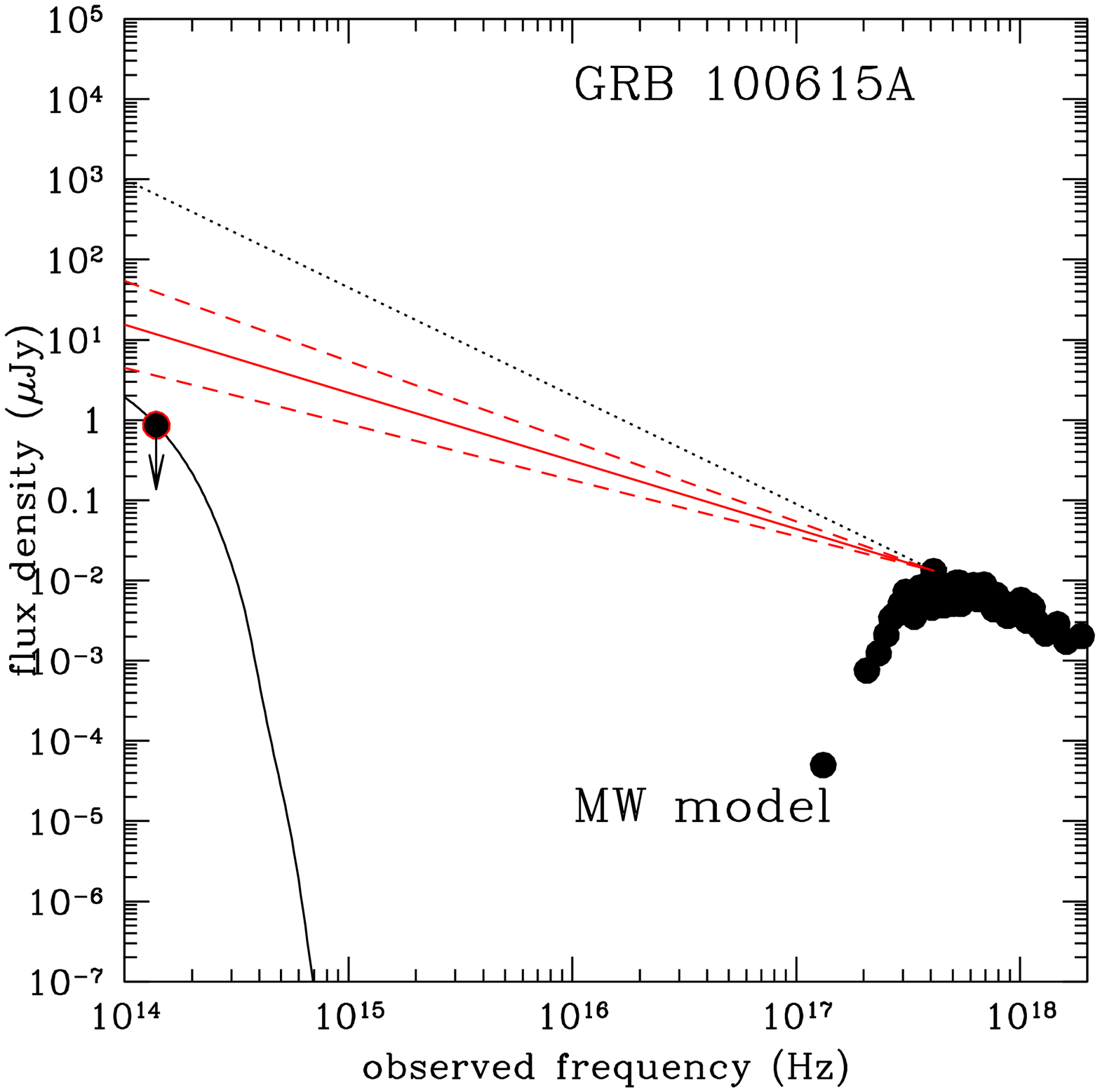}
\includegraphics[angle=-0,width=3.23cm]{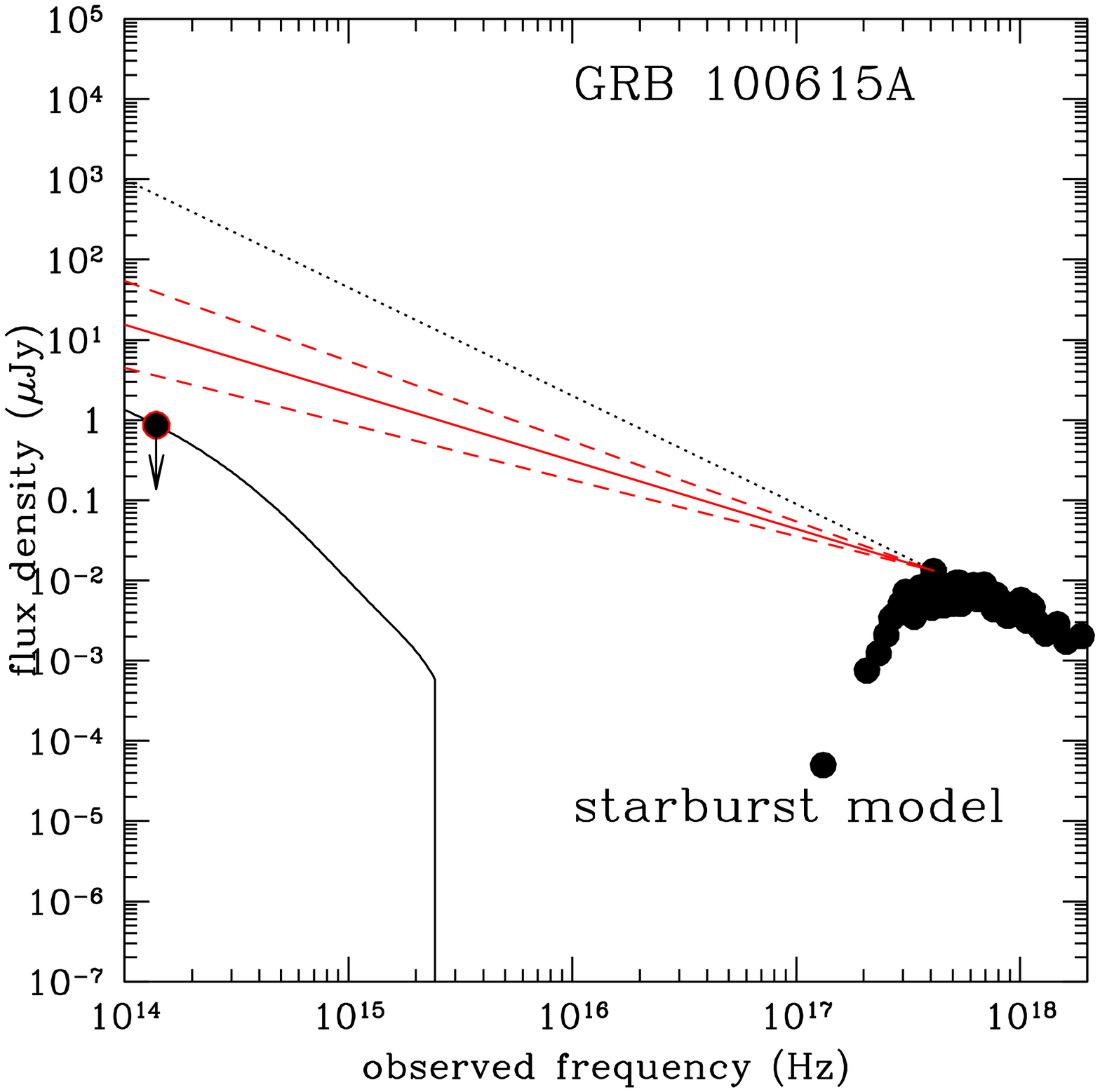}

\caption{GRB\,100615A SEDs at three different epochs: GROND $grizJHK$
  upper limits taken 24 minutes after the trigger (upper panels) and
  Gemini-North K and Ks upper limits taken 5.5 hours and 9.2 days
  after the trigger, respectively (middle and lower panels), with
  $\beta_{OX}$ fixed at $\beta_X-0.5$ and the optical suppression
  modelled with the MW or starburst extinction curves. Black dotted
  line represents the best-fit value for the X-ray spectral slope
  ($\beta_X=1.35$). Dashed red lines enclose the $90\%$ uncertainty in
  $\beta_{OX}$.}
\end{figure}

\section{Discussion}

We have analyzed two {\it Swift} `dark' GRBs, namely, GRB\,100614A,
and GRB\,100615A. These GRBs are dark according to every definition
proposed until now. They are not detected in the optical/NIR down to
very faint limits, despite follow-up campaigns at ground-based
facilities began minutes to hours from the BAT triggers (see Table
1). In addition, their optical-to-X-ray spectral indices satisfy
$\beta_{OX}<\beta_X-0.5$ (van der Horst et al. 2009 criterion).  The
identification of these two GRBs as dark bursts is the consequence of
their intense X-ray flux coupled to the optical/NIR missing
detections.

The outcome of our analysis is surprising. To explain the deepest NIR
upper limits (i.e. the less affected by dust) in terms of a flux
suppression described by either a MW or SMC dust extinction laws, $A_V
> 47$ ($A_V > 58$) mag is needed for GRB\,100614A (GRB\,100615A)
before one day and $A_V>22$ at 9 days after the trigger for
GRB\,100615A. Such extreme $A_V$ values have never been observed
before and require an explanation.
 
While a SMC-like extinction curve can adequately fit a large fraction
of the dust extinction from GRB host ISM, our present picture of GRB
host galaxies makes an extremely obscured environment of this kind a
very unlikely possibility. Indeed, GRBs, even reddened ones, are
hosted by blue or normal galaxies, which are commonly detected in
ordinary galaxy surveys. This favours a scenario of a host morphology
where the line of sight to the GRB is dusty, i.e., dust obscures only
localized regions (see e.g., Perley et al. 2009 and references
therein). An in situ obscuration appears to be insufficient to be
responsible for the extreme extinction levels we measure.

All these considerations hold if the optical radiation and X-rays are
part of the same synchrotron spectrum. They could originate from
different emission processes or even be produced in different,
independent emission regions. This is possible in particular during
the so-called "shallow phase", where X-ray emission may be dominated
by an emission component that differs from the one from which the
optical flux originates (e.g. Zhang 2007). Since the two SEDs of
GRB\,100614A, the first one of GRB\,100615A, and (marginally) the
second one of GRB\,100615A are all extracted during the shallow phases
of these GRBs, a different origin of the optical and X-ray emission in
these epochs could at least in part explain the optical darkness of
our GRBs. However, we note that for GRB\,100615A we have extracted a
SED at a very late time, about nine days after the end of the plateau
phase, and we have still obtained very high $A_V$ lower limits
($A_v>20$) assuming either a MW or a SMC extinction curve. These
values are less extreme than that obtained using the other SEDs, but
still very high, suggesting another or at least a concurring mechanism
to account for the GRB darkness.

This complementary explanation could be that local extinction recipes,
such as MW or SMC ones are inadequate for reproducing the optical
suppression in the host galaxies of these two GRBs. For example,
modelling the dust absorption using greyer extinction laws, such as
the attenuation curve obtained from the observations of starburst
galaxies proposed by Calzetti (1994), brings GRB\,100614A and
GRB\,100615A to require $A_V$ lower limits that are less extreme,
despite still being very high.

A mixture of moderate-to-high redshift and extinction can reduce the
dust level necessary to explain the SEDs. In Fig. 4, we plot as an
example the $A_V$ values as a function of $z$ obtained for the second
epoch of GRB\,100615A observations. Although the visual extinction is
considerably lower than for the $z=0$ case, $A_V \sim 10$ at $z=2$ and
$A_V \sim 4-5$ at $z=5$ is still required, regardless of the adopted
extinction recipe. Thus, an intrinsic origin and/or dust extinction,
coupled to a moderately high redshift could explain the darkness of
our GRBs.

\begin{figure}[t!]
\resizebox{\hsize}{!}{\includegraphics[clip=true]{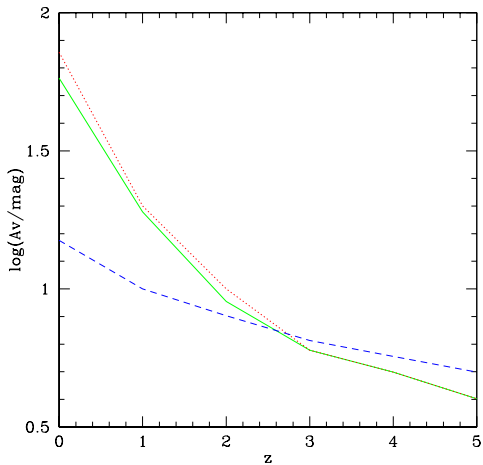}}
\caption{The visual extinction $A_V$ as a function of redshift for the
  second epoch SED of GRB\,100615A. Green solid, red dotted, and blue
  dashed lines are for the Milky Way, Small Magellanic Cloud, and
  starburst recipes, respectively.}
\label{spe1}
\end{figure}

A more exotic but intriguing possibility would be that these GRBs are
extremely high redshift events.  Assuming that the lack of any
detection in the reddest NIR band (K-band) is due to Ly$\alpha$
absorption from the intergalactic hydrogen neutral fraction, we can
set a redshift lower limit of $z>17$. The first population of very
massive stars (PopIII, $10^2M_\odot<M<10^3M_\odot$) is expected to
form at $z\sim 20$. Their death is supposed to leave behind black
holes of several tens of solar masses, which could be the early
progenitors of active galactic nuclei. These fast-spinning black holes
have a rotational energy of $\sim 10^{55}$ erg or more that can power
a GRB explosion. The isotropic energies of GRB\,100614A and
GRB\,100615A assuming $z=18$ are $1.3 \times 10^{54}$ erg and $7.2
\times 10^{53}$ erg, respectively (using the BAT fluences). However,
the reported isotropic energies must be considered as lower limits
both because of the conservative choice of the $z$ used and because
the BAT detector does not constrain the position of the peak emission
preventing a bolometric estimate of the emitted energy. The lack of
detection of any host galaxy candidate for these GRBs represents
additional support of this scenario.  On the other hand, against the
high redshift interpretation there is that the expected GRB rate at
$z>17$ is extremely low, between $0.5$ and $1$ GRB every $10$ yr
(Bromm \& Loeb 2006).

A possible way to differentiate between extremely high-redshift,
exotic extinction recipes and emission from distinct components for
sources such as these, would be to search for the afterglow in the
mid- or far-IR bands. A non-detection also in these bands could hardly
be explained using any extinction law and would definitely rule out a
high redshift origin. A multiband detection compatible with the X-ray
flux at late times (e.g. $> 10^5$ s after the trigger, i.e. after the
end of the "shallow phase") would instead favour the fireball
model. In this case, given the lack of NIR detections, the origin of
the darkness would be in exotic extinction or high redshift, depending
on the spectral shape in the mid- and far-IR bands.

\bibliographystyle{aa}

\end{document}